\documentclass[showpacs]{revtex4}
\usepackage{epsfig}

\newcommand{\beq}{\begin{equation}}
\newcommand{\eeq}{\end{equation}}
\newcommand{\beqa}{\begin{eqnarray}}
\newcommand{\eeqa}{\end{eqnarray}}
\def\ket#1{|\,#1\,\rangle}

\def\opone{\leavevmode\hbox{\small1\kern-3.8pt\normalsize1}}

\begin{document}
\title{Entanglement-assisted local transformations between inequivalent classes of three-particle entangled states}
\author{Iulia Ghiu $^{a,b}$}\email{iughiu@barutu.fizica.unibuc.ro}   \author{Mohamed Bourennane $^a$} \author{Anders Karlsson $^a$}  
\affiliation{ $^a$ Department of Microelectronics and Information Technology,
Royal Institute of Technology (KTH), Electrum 229,
 164 40 Kista, Sweden}
\affiliation{$^b$ Department of Physics,  University of Bucharest, P.O. Box MG-11, R-077125, Bucharest-M\u{a}gurele, Romania}

\date{\today}

\begin{abstract}
We show that bipartite and tripartite entangled states cannot be used as catalysis states to enable local transformations inbetween inequivalent classes of three-particle entangled states which are non-interchangeable under local transformations. We find the optimal protocol for conversion of a certain family of the W-states and an EPR pair into the GHZ state. 
\end{abstract}

\maketitle

Entanglement is the key physical resource in most quantum information processes, e.g. quantum teleportation \cite{Bennett}, two- or multiparty quantum cryptography\cite{Ekert,Buzek}, and quantum computation \cite{Shor,Grover}. Moving onwards from two-particle entangled states, much interest has been devoted to three-particle entangled states, notably the Greenberger-Horne-Zeilinger (GHZ) states \cite{Zeilinger I} 
\beq\label{GHZ}
\ket{GHZ}=\frac{1}{\sqrt 2}(\ket{000}+\ket{111}) 
\eeq
and their role in generalizations of Bell inequalities, as well as an enabling resource for quantum computation \cite{Comput}.
Whereas for two-particle states, entanglement local manipulation is known, see e.g. \cite{Jonathan,Nielsen2,Vidal2}, for three or multiparticle entanglement, there still remain unresolved issues concerning both entanglement transformations as well as
possible applications in quantum information processing.

 Recently, there has been
some interest devoted to so called W-states, an example being the state

\beq \label{w}\ket{W}={1\over \sqrt{3}}(\ket{001}+\ket{010}+\ket{100}).\eeq

An interesting property of this state is that if, say, particle one is traced out, there
remains a large degree of entanglement in particle two and three, or if
the state of the first particle is measured in the $ \{0,1 \}$ basis, then either
the state of particle two and three is maximally entangled, or in a product state. 

It was shown \cite{Vidal} that states of the GHZ-type and of the W-type
are inequivalent in the following sense: if we allow only Stochastic Local
quantum Operations and Classical Communications \cite{Bennett2}, abbreviated
SLOCC, then one cannot succeed in transforming states from the GHZ-class to
the W-class and vice-versa with a non-zero probability of success. We note
that
  in Refs.\cite{Cohen,Acin} the optimal distillation of the state GHZ
from one copy of an arbitrary tripartite entangled state has been presented.

Of some interest in entanglement transformations has been entanglement catalysis processes \cite{Plenio}. The questions we address here are: Can we perform the transformation between the two inequivalent classes using a catalysis state, such as shared Einstein-Podolsky-Rosen (EPR) pairs, quantum states of the form:
\beq 
\label{EPR}
\ket{ \Psi^+}=\frac{1}{\sqrt 2}(\ket{00} +\ket{11}),
\eeq
and which is the probability of success to transform W-states into GHZ-states and vice-versa by using SLOCC in presence of additional entanglement resources?

Our Letter is organised as follows: We introduce the two general classes of three-particle entangled states inequivalent under SLOCC. Then we review the one copy bipartite entanglement optimal manipulation and the concept of catalysis for bipartite states.
 We show that a catalysis procedure between W-states and GHZ-states using either a non-maximally entangled bipartite state, GHZ or W-states as catalysis states is not possible. Then we find the optimal protocol for conversion of a certain family of the W-states and an EPR pair into the state GHZ. 

Let us then first show, as an illustration, a simple way  to
convert the state W of Eq.(\ref{w}) into the state GHZ of Eq.(\ref{GHZ}) using a single EPR-pair. Suppose the three parties involved, Alice, Bob, and Charlie, share three-particle entanglement of the type
W, and at the same time, Alice and Bob share one EPR-pair. To transform the state, first Alice makes a measurement in the computational basis $\{0,1\}$. From Eq.(\ref{w}) we see that with propability $p = 2/3$, she projects out a two-particle maximally entangled state between Bob and Charlie,
and when that happens she also knows it with certainty.
  Now we have two EPR-pairs, one pair for Alice-Bob and another Bob-Charlie. Second,
Bob can prepare locally the state GHZ, then he teleports \cite{Bennett} the state of one particle to Alice and another one to Charlie by using the shared Alice-Bob and Bob-Charlie EPR-pairs, respectively \cite{Bennett2}. 

Moving onwards to the more general case, let us define the two classes of three-particle entangled states \cite{Vidal}, first the GHZ-class:
\beq \ket{ \psi_{GHZ}}=\sqrt K(c_\delta\ket{0}\ket{0}\ket{0}+s_\delta e^{i\phi} \ket{\varphi_A} \ket{\varphi_B} \ket{\varphi_C}),
\eeq
where $ \ket{\varphi_A}=c_\alpha\ket{0}+s_\alpha\ket{1},  \ket{\varphi_B}=c_\beta\ket{0}+s_\beta\ket{1},  \ket{\varphi_C}=c_\gamma\ket{0}+s_\gamma\ket{1}  $ and K is the normalization factor.

Second, the W-class is:
\beq \label{wclass}\ket{\psi_{W}}(a,b,c,d)=\sqrt a\ket{100}+\sqrt b\ket{010}+ \sqrt c\ket{001}+\sqrt d\ket{000},
\eeq
where $a,b,c>0, d\ge0$, and $ a+b+c+d=1$.
The state W of Eq. (\ref{w}) is characterized by: $d=0, a=b=c=1/3$.
It has been shown that there is not a local operator (invertible or non-invertible) $A\otimes B\otimes C$ (where $A,B$ and $C$ are the local operators of Alice, Bob, and Charlie, respectively) such that:
\beq \ket{\psi_W}=A\otimes B\otimes C \ket{\psi_{GHZ}},
\eeq
where $\ket{\psi_{GHZ}}$ and $\ket{\psi_W}$ belong to the GHZ-class and W-class, respectively \cite{Vidal}. 

Suppose now that two observers, Alice and Bob, share one copy of a pure bipartite entangled state $\ket{\Psi}_{AB}$ and that they would like to convert it into another pure bipartite entangled
state $\ket{\Phi}_{AB}$. The greatest probability of
success, if the two parties are allowed only to act by LOCC, is \cite{Vidal2}:
\beq \label{opt}
P(\Psi_{AB} \rightarrow \Phi_{AB}) = \min_{l\in [1,n]} \frac{\sum_{i=l}^n
\alpha_i}{\sum_{i=l}^n \beta_i},
\eeq
where $\alpha_i$ and $\beta_i$ are the Schmidt coefficients of $\Psi$ and
$\Phi$ defined as, 
\beqa
\ket{ \Psi}_{AB} = \sum_{i=1}^n \sqrt{\alpha_i} \ket{ i_A i_B}, \sum_{i=1}^n
\alpha_i = 1,  \nonumber \\
\ket{ \Phi}_{AB} = \sum_{i=1}^n \sqrt{\beta_i} \ket{ i_A i_B}, \sum_{i=1}^n
\beta_i = 1,
\eeqa
where $\ket{ i_A}$ and  $\ket{ i_B}$ are the bases for the quantum system $A$
and $B$.

In Ref.\cite{Plenio}, the catalysis transformation between two bipartite states $\ket{\psi_1}, \ket{\psi_2}$ is defined as follows: Suppose Alice and Bob share an entangled state $\ket{\psi_1}$, that cannot be converted into $\ket{\psi_2}$ by LOCC. A preparator can 'lend' them an entangled state $\ket{\phi}$. If the transformation
\beq 
\ket{\psi_1}\otimes \ket{\phi}\to \ket{\psi_2}\otimes \ket{\phi}
\eeq
is possible, then this protocol is called entanglement-assisted local transformation or catalysis transformation \cite{Plenio}, since the state $\ket{\phi}$ is not consumed. 

We start by studying the catalysis transformations between the two types of three-particle entangled states.

{\bf Proposition 1.}
{\it Transformations between W-states and GHZ-states cannot be catalyzed by a bipartite entangled  state.}

{\bf Proof.}

Let us consider that Alice, Bob, and Charlie share a W-state, and, at the same time, Bob and Charlie share a bipartite entangled state, the total state being:
\beqa \label{initial} 
&&\ket{\Phi}=(\sqrt a\ket{1}_A\ket{0}_B\ket{0}_C+\sqrt b\ket{0}_A\ket{1}_B\ket{0}_C+ \sqrt c\ket{0}_A\ket{0}_B\ket{1}_C)\nonumber\\
&&\otimes (\sqrt {\alpha}\ket{0}_B\ket{0}_C+ \sqrt {\beta}\ket{1}_B\ket{1}_C),
\eeqa
where $\alpha, \beta>0, \alpha +\beta =1$.
They would like to transform this state into the following one:
\beq\label{final}
\ket{\Psi}=\ket{ \psi_{GHZ}}_{ABC}\otimes (\sqrt {\alpha}\ket{0}_B\ket{0}_C+ \sqrt {\beta}\ket{1}_B\ket{1}_C). 
\eeq  

The ranks of the reduced density matrices of the above states (\ref{initial}), (\ref{final}) are:
\beq\label{rang}
r(\rho_A)=2, r(\rho_B)=4, r(\rho_C)=4.
\eeq
Consider that the state $\ket{\Phi}$ can be converted into $\ket{\Psi}$ by a noninvertible operator $A\otimes B\otimes C$:
\beq
\ket{\Psi}=A\otimes B\otimes C\ket{\Phi},
\eeq
that means at least one of the three operators do not have the maximal rank (2 for A, 4 for B and C). This would lead to a lower rank for the reduced density matrices. But this is not possible since the ranks of the three reduced density matrices of the initial and final states are identical and given by Eq.(\ref{rang}).

Consider now that the two states can be converted into each other by an invertible operator:
\beqa\label{abcst}
A\otimes B\otimes C\ket{\Phi}&=&
\sqrt{a \alpha }(A\ket{1})(B\ket{00})(C\ket{00})+ \sqrt{a \beta}(A\ket{1})(B\ket{01})(C\ket{01})\nonumber \\
&+&\sqrt{b \alpha}(A\ket{0})(B\ket{10})(C\ket{00})+\sqrt{b \beta}(A\ket{0})(B\ket{11})(C\ket{01})\nonumber\\
&+&\sqrt{c \alpha}(A\ket{0})(B\ket{00})(C\ket{10}+ \sqrt{c \beta}(A\ket{0})(B\ket{10})(C\ket{11}).
\eeqa
Since the three operators are invertible, we obtain $A\ket{0},A\ket{1} $ are linearly independent vectors; $B\ket{00}, B\ket{01}, B\ket{10}, B\ket{11}$ are independent, and the same for  $C\ket{00}, C\ket{01}, C\ket{10}, C\ket{11}$.  That means the minimal number of product terms for the state (\ref{abcst}) is constant under invertible SLOCC \cite{Vidal} and equal to 6, which is different of the minimal number of product states for $\ket{\Psi}$ (which is 4). Hence it is not possible to use bipartite states as catalysis states.

{\bf Observation}. Following the same arguments as before, we can prove that there is no catalysis using two non-maximally bipartite states shared by Bob and Charlie:
\beq 
\ket{\psi_W}_{ABC}\otimes \ket{\psi_1}_{BC}\otimes \ket{\psi_2}_{BC}\not\to \ket{\psi_{GHZ}}\otimes\ket{\psi_1}_{BC}\otimes \ket{\psi_2}_{BC}. 
\eeq

{\bf Proposition 2.}
{\it Transformations between W-states and GHZ-states cannot be catalyzed by a tripartite entangled  state.}

{\bf Proof.}

We have to prove that the following transformations are not possible:
\beq \label{stare1}
\ket{\psi_W}\otimes \ket{\psi_W}\not\to \ket{\psi_{GHZ}}\otimes \ket{\psi_W}; 
\eeq
\beq\label{stare2}
\ket{\psi_W}\otimes \ket{\psi_{GHZ}}\not\to \ket{\psi_{GHZ}}\otimes \ket{\psi_{GHZ}}.
\eeq

Suppose that there are local protocols that can perform the above transformations. Since the ranks of the reduced density matrices of the initial and final states (\ref{stare1}), (\ref{stare2}) are:
$r(\rho_A)=4, r(\rho_B)=4, r(\rho_C)=4$, that means that only an invertible local operator would perform the transformations. But the minimal number of product states in the initial and final states in (\ref{stare1}) and (\ref{stare2}) is different and this leads to contradiction since the minimal number of product states has to be constant under invertible local operators SLOCC \cite{Vidal}.

Suppose now we want to transform the state $\ket{\Phi}$ given by Eq.(\ref{initial}) into the state GHZ. Since the catalysis is not possible, the most general final state obtained by SLOCC is:
\beq\label{stf}
\ket{\chi}=\frac{1}{\sqrt 2}(\ket{000}+\ket{111})\otimes \ket{00}
\eeq
(or a local unitary equivalent one).

{\bf Proposition 3.}
{\it Suppose Alice, Bob, and Charlie share a copy of the following W-state (\ref{wclass}) (the three particles are denoted by the indices "1", "2", and "3"):
\beq 
\ket{\psi_{W}}_{123}(a,a,1-2a,0)=(\sqrt a\ket{100}+\sqrt a\ket{010}+ \sqrt{1-2a}\ket{001})_{123},
\eeq
where $a\in \left[\frac{1}{3},\frac{1}{2}\right[$.
 Consider that a preparator, Daniel, can send to two of them an EPR pair. They want to transform this state into the state GHZ (\ref{GHZ}) with the help of the EPR pair ``45''(see Fig. 1):
\beq \label{transf1}
\ket{\psi_W}_{123}(a,a,1-2a,0)\otimes \ket{EPR}_{45}\to \ket{GHZ}_{125}\otimes \ket{00}_{34}.
\eeq

Then the highest probability is equal to '$2a$' and is obtained when the preparator sends 
the EPR pair ``45'' to Alice and Charlie, or to Bob and Charlie. The optimal protocol for achieving
 the transformation consists in three steps: \\
(a) Charlie measures his particle ``3'' in the computational basis. After this measurement, he has to send the outcome to Alice and Bob. \\
 Let us consider that Bob and Charlie share the EPR pair.\\
(b) Bob applies a CNOT operation onto his particles, where the particle ``2'' represents the source, while particle ``4'' is the target.\\
(c) Bob measures particle ``4'' in the computational basis.}

{\bf Proof.}

The initial state can be written as follows:
\beqa
&&\ket{\phi}=\ket{\psi_W}_{123}\otimes \ket{EPR}_{45}=(\sqrt a\ket{100}+\sqrt a\ket{010}\nonumber\\
&&+\sqrt{1-2a}\ket{001})_{123}\otimes \frac{1}{\sqrt 2}(\ket{00}+ \ket{11})_{45}; 
\eeqa

First, we will compute the probability given by the above protocol, and second, we will prove that this protocol is the optimal one for the transformation given by Eq.(\ref{transf1}).

Step (a): After the measurement, Charlie will obtain the state $\ket{0}_3$ with the probability $2a$, and the state of the whole system will be projected onto:
\beqa
\ket{\phi^1}&=&\frac{1}{2}\left( \ket{1}_1\ket{00}_{24}\ket{0}_5+\ket{1}_1\ket{01}_{24}\ket{1}_5\right)+\nonumber\\
&&+\frac{1}{2}\left(\ket{0}_1\ket{10}_{24}\ket{0}_5+\ket{0}_1\ket{11}_{24}\ket{1}_5 \right).
\eeqa

\begin{figure} 
\includegraphics[height=7cm,width=14cm]{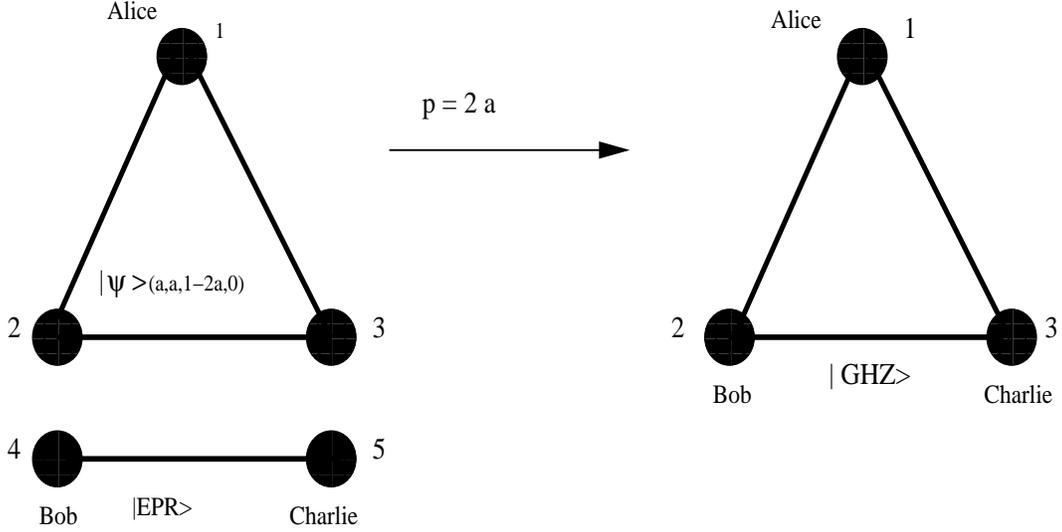}
\caption{Schematic of the transformation from $\ket{\psi_W}(a,a,1-2a,0)$ into the state GHZ using one EPR-pair
(the thick lines represent the entanglement).}
\end{figure}

Step (b): The CNOT gate performed by Bob will lead to the following state:
\beqa
\ket{\phi^2}&=&\frac{1}{2}\left( \ket{1}_1\ket{0}_2\ket{0}_5+\ket{0}_1\ket{1}_2\ket{1}_5\right)\ket{0}_4+\nonumber\\
&&+\frac{1}{2}\left(\ket{1}_1\ket{0}_2\ket{1}_5+\ket{0}_1\ket{1}_2\ket{0}_5\right)\ket{1}_4 .
\eeqa

Step (c): Regardless of Bob's outcome measurement of particle ``4'', the final state is local unitary equivalent to the state GHZ. We have obtained the final state with the total probability:
\beq \label{2a} P=2a.
\eeq 

In the Appendix it is proved that the above local protocol is optimal.

We can find a similar optimal protocol for conversion of  the following states:
\beqa
&&(\sqrt{1-2b}\ket{100}+\sqrt b\ket{010}+\sqrt b\ket{001})_{123}\otimes \frac{1}{\sqrt 2}(\ket{00}+ \ket{11})_{45};\\
&&(\sqrt{c}\ket{100}+\sqrt{1-2c}\ket{010}+\sqrt c\ket{001})_{123}\otimes \frac{1}{\sqrt 2}(\ket{00}+ \ket{11})_{45}, 
\eeqa
where  $b, c\in \left[\frac{1}{3},\frac{1}{2}\right[$,
into the state of Eq.(\ref{stf}).

Let us discuss how we can apply the above procedure backwards to go from
the GHZ-class to the W-class. Suppose that Alice, Bob, and Charlie share a GHZ-state, and at the same time Alice and Bob share an EPR-pair. One can obtain an EPR-pair shared by Bob and Charlie from the GHZ-state \cite{Linden} with probability one if Alice performs a measurement in x-basis ($\ket{0'}, \ket{1'}$):
\beq
\ket{GHZ}=\frac{1}{\sqrt 2}\left[ \ket{0'}_A\frac{1}{\sqrt 2}(\ket{00}+\ket{11})_{BC}+\ket{1'}_A\frac{1}{\sqrt 2}(\ket{00}-\ket{11})_{BC}\right].
\eeq
Bob can prepare locally an arbitrary three-particle entangled state, and by using the two EPR-pairs, the three observers will share the three-particle state.

In conclusion, we have shown that catalysis between the two inequivalent classes of W and GHZ-states under local operations is not possible. Then, we have demonstated how stochastic local operations followed by classical communications assisted by additional entanglement can be used to perform transformations between two states of the two inequivalent classes.
 We believe these studies will be of interest to
get insights into multi-particle entanglement manipulations with applications.

\section*{Acknowledgments}
This work was supported by the
European IST QuComm project, the Swedish Technical Science Research
Council (TFR), the Swedish Natural Science Research
Council (NFR), and the Swedish Institute (SI) through a visiting researcher fellowship. A. Karlsson also thanks the Erwin Schr\" odinger Institute, 
Vienna, for support during the quantum information program.

\appendix
\section{The optimal transformation (Proposition 3)}
\renewcommand\theequation{\thesection.\arabic{equation}}
\setcounter{equation}{0} 

Let us suppose that Alice, Bob, and Charlie share a copy of the following W-state (the three particles are denoted by the indices "1", "2", and "3"):
\beq 
\ket{\psi_{W}}_{123}=(\sqrt a\ket{100}+\sqrt a\ket{010}+ \sqrt{1-2a}\ket{001})_{123},
\eeq
where $a\in \left[\frac{1}{3},\frac{1}{2}\right[$, and, at the same time, Bob and Charlie have an EPR pair, the total state being:
\beqa
&&\ket{\phi}=\ket{\psi_W}_{123}\otimes \frac{1}{\sqrt 2}(\ket{00}+\ket{11})_{45}\nonumber\\
&&=\sqrt{\frac{a}{2}}\ket{1}_1\ket{00}_{24}\ket{00}_{35}+\sqrt{\frac{a}{2}}\ket{1}_1\ket{01}_{24}\ket{01}_{35}\nonumber\\
&&+\sqrt{\frac{a}{2}}\ket{0}_1\ket{10}_{24}\ket{00}_{35}+\sqrt{\frac{a}{2}}\ket{0}_1\ket{11}_{24}\ket{01}_{35}\nonumber\\
&&+\sqrt{\frac{1-2a}{2}}\ket{0}_1\ket{00}_{24}\ket{10}_{35}+\sqrt{\frac{1-2a}{2}}\ket{0}_1\ket{01}_{24}\ket{11}_{35}.
\eeqa
 They want to perform the transformation given by Eq.(\ref{transf1}):
\beq \label{transf}
\ket{\phi} \to \ket{\chi},
\eeq
where $\ket{\chi}=\frac{1}{\sqrt 2}\ket{0}_1\ket{00}_{24}\ket{00}_{35}+\frac{1}{\sqrt 2}\ket{1}_1\ket{10}_{24}\ket{01}_{35}$.

Let us now review the concept of bipartite splitting \cite{DurCir}: consider N observers who share N particles. A partition of the N particles into two groups is called a bipartite splitting if the observers who have the particles in the same group can act jointly on them. In the following we will consider the three posibilities of bipartite splitting of the particles in the initial and final state. Then, using the existence of a Schmidt decomposition of a bipartite system \cite{Nielsen,Ekert2}, we find an upper bound for the probability of local conversion.

Case (i): Bipartition (Alice, Bob)-Charlie. The initial state can be written in the Schmidt decomposition:
\beqa
&&\ket{\phi}=\sqrt a\ket{\bar{0}}_{AB}\ket{\bar{0}}_C+\sqrt a\ket{\bar{1}}_{AB}\ket{\bar{1}}_C\nonumber\\
&&+\sqrt{\frac{1-2a}{2}}\ket{\bar{2}}_{AB}\ket{\bar{2}}_C+\sqrt{\frac{1-2a}{2}}\ket{\bar{3}}_{AB}\ket{\bar{3}}_C,
\eeqa
where $\ket{\bar{0}}_{AB}=\frac{1}{\sqrt 2}(\ket{1}\ket{00}+\ket{0}\ket{10})$,$\ket{\bar{1}}_{AB}=\frac{1}{\sqrt 2}(\ket{1}\ket{01}+\ket{0}\ket{11})$, $\ket{\bar{2}}_{AB}=\ket{0}\ket{00}$, $\ket{\bar{3}}_{AB}=\ket{0}\ket{01}$, $\ket{\bar{0}}_C=\ket{00}$, $\ket{\bar{1}}_C=\ket{01}$, $\ket{\bar{2}}_C=\ket{10}$, $\ket{\bar{3}}_C=\ket{11}$.

The Schmidt decomposition of the final state is:
\beq
\ket{\chi}=\frac{1}{\sqrt 2}\ket{0'}_{AB}\ket{\bar{0}}_C+\frac{1}{\sqrt 2}\ket{1'}_{AB}\ket{\bar{1}}_C,
\eeq
with $\ket{0'}_{AB}=\ket{0}\ket{00}$ and $\ket{1'}_{AB}=\ket{1}\ket{10}$. We will denote by $\psi_1\sim \psi_2$ two states $\psi_1,\psi_2$ which are local unitary equivalent. We have:
\beq
\ket{\chi}\sim \ket{\epsilon}=\frac{1}{\sqrt 2}\ket{\bar{0}}_{AB}\ket{\bar{0}}_C+\frac{1}{\sqrt 2}\ket{\bar{1}}_{AB}\ket{\bar{1}}_C.
\eeq

The maximal probability of success to transform $\ket{\phi}\to \ket{\epsilon}$ is given by Eq.(\ref{opt}):
$P_1=1$.

Case (ii): Bipartition Alice-(Bob,Charlie). The Schmidt decomposition is:
\beq  \label{statei}
\ket{\phi}=\sqrt{1-a}\ket{\bar{0}}_A\ket{\bar{0}}_{BC}+\sqrt a\ket{\bar{1}}_A\ket{\bar{1}}_{BC},
\eeq
where $\ket{\bar{0}}_A=\ket{0}$, $\ket{\bar{1}}_A=\ket{1}$, $\ket{\bar{0}}_{BC}=\frac{1}{\sqrt{1-a}}(\sqrt{\frac{a}{2}}\ket{10}\ket{00}+\sqrt{\frac{a}{2}}\ket{11}\ket{01}+\sqrt{\frac{1-2a}{2}}\ket{00}\ket{10}+\sqrt{\frac{1-2a}{2}}\ket{01}\ket{11}) $, $\ket{\bar{1}}_{BC}=\frac{1}{\sqrt 2}(\ket{00}\ket{00}+\ket{01}\ket{01})$.

The final state is:
\beq
\ket{\chi}=\frac{1}{\sqrt 2}\ket{\bar{0}}_A\ket{0'}_{BC}+\frac{1}{\sqrt 2}\ket{\bar{1}}_A\ket{1'}_{BC}\sim \ket{\mu}=\frac{1}{\sqrt 2}\ket{\bar{0}}_A\ket{\bar{0}}_{BC}+\frac{1}{\sqrt 2}\ket{\bar{1}}_A\ket{\bar{1}}_{BC},
\eeq

The maximal probability to transform $\ket{\phi}$ into $\ket{\mu}$ is given by Eq.(\ref{opt}):
\beq
P_2=2a.
\eeq

Case (iii) Bipartition (Alice,Charlie)-Bob. The Schmidt decomposition of the initial state is:
\beqa
&&\ket{\phi}=\sqrt{\frac{1-a}{2}}\ket{\bar{0}}_{AC}\ket{\bar{0}}_B+\sqrt{\frac{1-a}{2}}\ket{\bar{1}}_{AC}\ket{\bar{1}}_B\nonumber\\
&&+\sqrt{\frac{a}{2}}\ket{\bar{2}}_{AC}\ket{\bar{2}}_B+\sqrt{\frac{a}{2}}\ket{\bar{3}}_{AC}\ket{\bar{3}}_B,
\eeqa
where $\ket{\bar{0}}_{AC}=\sqrt{\frac{a}{1-a}}\ket{1}\ket{00}+\sqrt{\frac{1-2a}{1-a}}\ket{0}\ket{10}$, $\ket{\bar{1}}_{AC}=\sqrt{\frac{a}{1-a}}\ket{1}\ket{01}+\sqrt{\frac{1-2a}{1-a}}\ket{0}\ket{11}$, $\ket{\bar{2}}_{AC}=\ket{0}\ket{00}$, $\ket{\bar{3}}_{AC}=\ket{0}\ket{01}$,  $\ket{\bar{0}}_B=\ket{00}$, $\ket{\bar{1}}_B=\ket{01}$, $\ket{\bar{2}}_B=\ket{10}$, $\ket{\bar{3}}_B=\ket{11}$.

We have in this case
$P_3=1$.

We will show that the maximal probability $P_{max}(LOCC)$ to convert $\ket{\phi}$ into $\ket{\chi}$ by LOCC satisfies the following condition:
\beq \label{bound}
P_{max}(LOCC)\le \hspace{0.1cm} \mbox{min} \{P_i\}=2a, 
\eeq
where $P_i$, i=1,2,3 are the probabilities for conversion of the bipartite states presented in cases (i),(ii),(iii), respectively.

Suppose that there is a protocol $\cal{P}$ for this conversion by LOCC with a probability of success P(LOCC) such that:
\beq
P(LOCC)>P_j=\mbox{min} \{P_i\}.
\eeq
Then the observers who share the bipartite state can use the protocol $\cal{P}$ in order to perform the transformation, which would give a higher probability than $P_j$. However this probability, obtained using the Eq.(\ref{opt}), is the maximal probability for conversion of the two bipartite states and this leads to contradiction; that means the Eq.(\ref{bound}) is satisfied.

The probability given by the local protocol presented in Proposition 3 is equal to the upper bound given by Eq.(\ref{bound}) and equal to $2a$, that means the local protocol is optimal. In Ref. \cite{Ghiu} a generalization of Eq.(\ref{bound}) for N-particle states is presented.


\vspace{-0.5cm}

\begin{thebibliography}{99}

\bibitem{Bennett} C.H. Bennett, G. Brassard, C. Crepeau, R. Jozsa, A. Peres,
and W.K. Wooters, Phys. Rev. Lett. {\bf 70}, 1895 (1993).
\bibitem{Ekert} A. K. Ekert, Phys. Rev. Lett. {\bf 67}, 661 (1991).
\bibitem{Buzek} M. Hillary, V. Buzek and A. Bertaiume,  Phys. Rev. A {\bf 59}, 1829 (1999).
\bibitem{Shor} P.W. Shor in {\em Proceedings of the Symposium on the
Foundation of Computer Science, 1994, Los Alamonitos, California} (IEEE
Computer Society Press,
New York  124 (1994).

\bibitem{Grover} L.K. Grover, Phys. Rev. Lett. {\bf 79},  325 (1997).

 \bibitem{Zeilinger I} D. M. Greenberger, M.A. Horne, A. Shimony, and A.
Zeilinger,   Am. J. Phys., {\bf 58}
 1131 (1990).


\bibitem{Comput}L.K. Grover, quant-ph/9704012; J. I. Cirac, A. K. Ekert, S.F. Huelga, C. Macchiavello, Phys. Rev. A {\bf 59}, 4249 (1999);   D. Gottesman and I.L. Chuang,  Nature {\bf 402},  390  (1999); D. Collins, N. Linden, S. Popescu, quant-ph/0005102 ; J. Eisert, K. Jacobs, P. Papadopoulos, M. B. Plenio, Phys. Rev. A {\bf 62}, 052317 (2000).


\bibitem{Jonathan}D. Jonathan and M. Plenio, Phys. Rev. Lett. {\bf 83}, 1455 (1999).

\bibitem{Nielsen2}M. A. Nielsen, Phys. Rev. Lett. {\bf 83}, 436 (1999).  
\bibitem{Vidal2} G. Vidal, Phys. Rev. Lett. {\bf 83}, 1046 (1999).

\bibitem{Vidal}W. D\"{u}r, G. Vidal, and J. I. Cirac, Phys. Rev. A {\bf 62}, 062314 (2000). 


 \bibitem{Bennett2}C.H. Bennett, S. Popescu, D. Rohrlich, J.A. Smolin, and A.V. Thapiliyal, Phys. Rev. A {\bf 63}, 012307 (2001). 

 \bibitem{Plenio}D. Jonathan and M. Plenio, 
Phys. Rev. Lett. {\bf 83}, 3566 (1999).
\bibitem{Linden}N. Linden, S. Popescu, B. Schumacher, and M. Westmoreland, quant-ph/9912039 (1999). 
\bibitem{Nielsen}M. A. Nielsen, quant-ph/0011036.

\bibitem{Ekert2}A. Ekert and P. L. Knight, Am. J. Phys. {\bf 63}, 415 (1993).
\bibitem{Cohen}O. Cohen, T. A. Brun, Phys. Rev. Lett. {\bf 84}, 5908 (2000).
\bibitem{Acin}A. Ac\'{\i}n, E. Jan\'{e}, W. D\"{u}r, and G. Vidal, Phys. Rev. Lett. {\bf 85}, 4811 (2000). 
\bibitem{DurCir}W. D\"{u}r and J. I. Cirac, Phys. Rev. A {\bf 61}, 042314 (2000).
\bibitem{Ghiu} I. Ghiu, M. Bourennane, and A. Karlsson, Rom. Rep. Phys. {\bf 57}, 283 (2005).


\end{thebibliography}
\end{document}